\documentclass[conference]{IEEEtran}
\IEEEoverridecommandlockouts
\usepackage{cite}
\usepackage{amsmath,amssymb,amsfonts}
\usepackage{lipsum}
\usepackage{algorithmic}
\usepackage{graphicx}
\usepackage{textcomp}
\usepackage{comment}
\usepackage{xcolor}
\usepackage{multirow}
\usepackage{booktabs}
\usepackage{makecell}
\usepackage{subcaption}
\usepackage{gensymb}

\def\BibTeX{{\rm B\kern-.05em{\sc i\kern-.025em b}\kern-.08em
    T\kern-.1667em\lower.7ex\hbox{E}\kern-.125emX}}

\begin{document}

\title{Three-phase model of unbalanced distribution networks with DERs}

\author{
\IEEEauthorblockN{S. Perna, C. Lillo, A.R. Di Fazio, M. Russo, G.M. Casolino, P. Varilone, P. Verde}
\IEEEauthorblockA{\textit{DIEI - Universit\`{a} degli Studi di Cassino e del Lazio Meridionale}\\
Cassino, Italy \\
\{sara.perna, a.difazio, russo, casolino, varilone, verde\}@unicas.it, carmine.lillo@studentmail.unicas.it}
}
\maketitle

\begin{abstract}
 Classical DistFlow equations for steady-state distribution network analysis fail to capture the inherent imbalances of three-phase systems arising from asymmetrical lines, loads, and distributed energy resources (DERs). This paper extends the classical power flow (PF) equations into a rigorous, non-approximated three-phase formulation, termed Dist3Flow. The proposed branch flow model (BFM) utilizes the real and imaginary components of nodal voltages and the active and reactive power flows as state variables. Lines are modelled by nonlinear forward and backward equations, while loads and DERs are represented via ZIP models and P–Q control, respectively. By incorporating specific boundary conditions at the terminal nodes, the formulation generalizes PF analysis to both radial and closed-ring topologies. The solution is obtained by using a backward/borward sweep (BFS) algorithm. The approach is validated against OpenDSS across various configurations, considering open-ring and closed-ring topologies with and without DERs.

\end{abstract}


\section*{Nomenclature}

For the generic three-phase quantity $A$ (i.e., voltage $V$ and current $I$ of node \textit{n}), the following notation stands\vspace{-3pt}
$$ \mathbf{\bar A}_n=\left[ {\bar A}_n^a\, {\bar A}_n^b\, {\bar A}_n^c\,\right]^T\hspace{-2pt}\, \mathrm{with} \,\, {\bar A}_n^i = A_{n,R}^i +\mathrm{j\,}A_{n,I}^i =A_n^i e^{\left(\mathrm{j}\,\angle{{\bar A}_{n}^i}\right)}\vspace{-3pt}$$
where $i\in \{a,b,c\}$ and superscript $T$ representing the simple transposition operator (not conjugate transposition). 
Moreover, $\mathbf{{\check A}}_n$ represents the conjugate of $\mathbf{{\bar A}}_n$, with the generic element ${\check  A}_n^i$ being the conjugate of ${\bar  A}_n^i$.

For apparent power, it is assumed \vspace{-6pt}
$$ \mathbf{\dot S}_n= \mathbf{P}_n\!\!+\mathrm{j\,}\mathbf{Q}_n\!=\!\left[ {\dot S}_n^a \,\, {\dot S}_n^b \,\, {\dot S}_n^c\,\right]^T\!\! \quad \mathrm{with}\,\, {\dot S}_n^i = P_{n}^i +\mathrm{j\,}Q_{n}^i\vspace{-3pt}$$

The impedance of the line between node $n-1$ and node $n$ is represented by matrix $\mathbf{\dot Z}_{n-1,n}$; its generic element ${\dot Z}_{n-1,n}^{ij}$, representing the coupling between phase $i$ and phase $j$, is\vspace{-3pt}
$${\dot Z}_{n-1,n}^{ij} = r_{n-1,n}^{ij} +\mathrm{j\,} x_{n-1,n}^{ij}\quad \mathrm{with} \,\, i,j\in \{a,b,c\} $$
Finally, the operator $\odot$ represents the Hadamard (or Schur) product among vectors or matrices.

\section{Introduction}
Power flow (PF) analysis is a fundamental tool for steady-state power system assessment. Its primary objective is to determine the network's electrical state—specifically nodal voltage magnitudes, angles, and branch PFs. Accurate PF results are essential for system operation, expansion planning, and grid management~\cite{Wang_2025, Babiker_2025}, while also providing critical support for electricity market operations.


Classical PF methods, originally designed for symmetric and balanced transmission networks, face significant limitations in distribution systems. Unlike transmission grids, that are modeled under the assumption of symmetric and balanced networks by adopting  single-phase equivalent, MV and LV distribution systems are inherently unbalanced due to structural asymmetries in the lines and imbalanced loads. Moreover, the rapid spread of distributed energy resources (DERs) — including DGs, storage, and electric vehicles — is increasing the complexity of system modeling, requiring three-phase PF models for unbalanced distribution grids with DERs~\cite{Neiva_2021, Singh_2023}. Furthermore, classical PF solving algorithms for transmission systems may prove inadequate when applied to radial or weakly-meshed topologies of distribution networks, which are characterized by lines with high r/x ratio.


The main methods proposed in the literature for solving PF in unbalanced three-phase distribution systems can be classified into three distinct categories, according to their mathematical formulations and solution algorithms.

The \textit{backward/forward sweep} (BFS) method was originally proposed for radial, symmetric, and balanced distribution systems in~\cite{BaranWu_1989}. It adopts a branch flow model (BFM) governed by the DistFlow equations; the solving algorithm is derived from the ladder iterative method~\cite{Kersting}, which consists of two alternating stages: a backward sweep, in which branch currents are aggregated, and a forward sweep, in which nodal voltages are updated. To derive three-phase formulations, BFMs have been proposed based on current flow representations~\cite{Kersting}. In this three-phase framework, the compensation method has also been applied to use BFS in weakly meshed distribution networks~\cite{Shirmohammadi_1988} and in the presence of DGs with PV control~\cite{Cheng_1995}. In~\cite{Khushalani_2007} the BFS has been extended to include DGs that dynamically switch between PV and PQ control modes during iteration. Computational efficiency is enanched in~\cite{Chang_2007} by decomposing the forward sweep into independent resistive networks for real and imaginary voltage components, utilizing a linear proportionality principle to update downstream voltages.

The \textit{Newton–Raphson} (NR) methods adopt a nodal current injection model for loads and DGs while lines are represented through three-phase impedances and admittances. The NR algorithm iteratively corrects the mismatch in current injections using the Jacobian matrix. In this framework, the main problem is the computational efficiency of the Jacobian matrix inversion. In~\cite{Teng_2002}, exploiting radial topology, an upper triangular Jacobian matrix is obtained, so as to solve PF through a simple substitution method, avoiding LU factorization employed in classical NR approaches. In~\cite{Garcia_2000}, the three-phase current injections are modelled in rectangular coordinates, resulting in a sparse Jacobian matrix that preserves the block structure of the original nodal admittance matrix.

The \textit{Gauss} or \textit{Zbus} methods rely on the overall three-phase nodal admittance matrix of the network to which LU factorization is typically applied to obtain the Zbus matrix. Bus voltage deviations are determined via Zbus matrix by applying the principle of superposition to nodal current injections. Then, current injections are updated using the newly calculated voltages. The process is iterated according to the Gauss algorithm~\cite{Chen_1991}. To reduce the computational burden due to LU factorization, a graph-theory-based method is proposed in~\cite{Yang_2016}, which exploits the sparsity of the nodal admittance matrix to directly obtain the Zbus matrix. In~\cite{Yang_2022} a Zbus-based approach is combined with a highly accurate initial estimate of nodal voltages, to reduce the number of iterations or, in some cases, obtain the solution in a single step.

This paper adopts a BFS approach based on BFM, extending the DistFlow equations from~\cite{BaranWu_1989} to a three-phase formulation, called Dist3Flow. Actually, three-phase extensions of the DistFlow equations have already been proposed in literature. However, they frequently rely on simplifying assumptions: \cite{Sankur_2016} assumes that the ratio of phase-voltage amplitudes is constant, the three-phase phasors are equally spaced, and losses are neglected; similarly, \cite{Chen_2018} neglects mutual impedances and network losses while \cite{Lin_2024} includes shunt components but assumes uniform current phasor angles, linearizes load models, and omits squared current terms in voltage drop calculations. A detailed analysis of the inaccuracy introduced by the approximations introduced by various methods is reported in~\cite{Inaolaji_2021}.


This paper presents an exact Dist3Flow formulation that avoids simplifying assumptions to ensure high accuracy. The model uses as state variables the active and reactive PFs in each branch, and the real and imaginary components of the sending-end voltages. Distribution lines are described through forward and backward nonlinear equations, expressing variables at the receiving node as functions of those at the sending node and vice versa. Loads are represented by ZIP model and DERs by assuming P-Q control.
The PF problem is solved using the BFS algorithm, which alternates between forward and backward sweeps at each iteration by imposing the boundary conditions at the slack bus (i.e., fixed real and imaginary  components of the voltage), and the ending node of the feeder (i.e., null outflowing active and reactive powers). The formulation is also extended to closed-ring topologies. Validation is carried out through a case study, with results benchmarked against OpenDSS.\vspace{-6pt}

\section{Modelling}

The distribution network consists of uncontrolled loads and DERs, modeled for simplicity as a single main feeder with $N$ nodes and $N-1$ branches, operating either radially or in a closed-loop configuration; however, the model can be extended to include laterals.\vspace{-6pt}

\subsection{DER and load models}
Concerning DERs, they typically inject assigned values of active and reactive powers (indicated with superscript $*$)
\vspace{-3pt}
\begin{equation}\label{eq:DER}
    \mathbf{\dot S}_{n}^{DER*}= \mathbf{P}_n^{DER*}+\mathrm{j\,}\mathbf{Q}_n^{DER*}
    \vspace{-3pt}
\end{equation}
Concerning uncontrolled loads, they can be modelled according to the widely-used ZIP model
\begin{equation}\label{eq:load}
    \mathbf{\dot S}_{n}^L= \mathbf{\dot S}_n^{L*}+ \mathbf{\overline V}_n \odot \mathbf{\check I}_n^{L*}+\mathbf{\overline V}_n \odot \mathbf{\check Y}_n^{L*} \odot \mathbf{\check V}_n
    \vspace{-3pt}
\end{equation}
where starred symbols always refer to assigned quantities.\vspace{-6pt}

\subsection{Line model}
\begin{figure}
    \centering
    \includegraphics[width=0.9\linewidth]{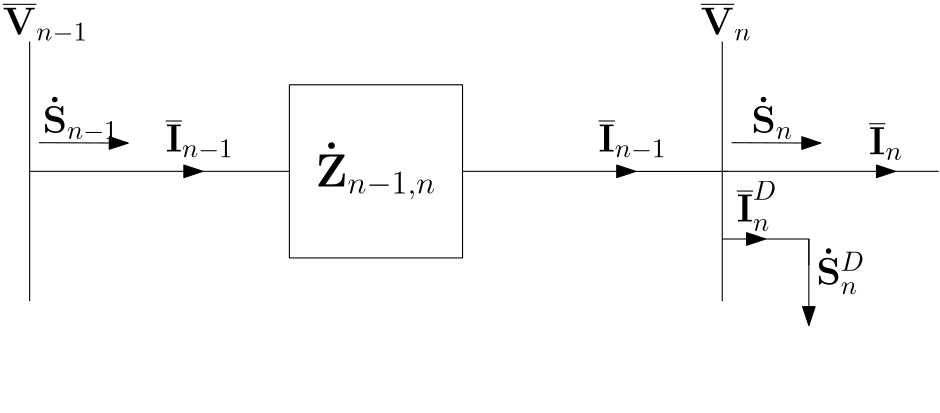}\vspace{-12pt}
    \caption{Three-phase line representation\vspace{-6pt}}
    \label{fig:linea}\vspace{-6pt}
\end{figure}

The generic line connecting node $n-1$ to node $n$ in Fig.~\ref{fig:linea} is modelled by a $3\times 3$ matrix $\mathbf{\dot Z}_{n-1,n}$ (i.e., Kron reduction~\cite{Kersting}). Line shunt components are neglected for simplicity; however, the model can be extended to include them. Uncontrolled loads and DERs connected at node $n$ are jointly modeled by the net power demand, denoted by the superscript $D$.

Writing Kirchhoff’s voltage and current laws, respectively, yields\vspace{-3pt}
\begin{eqnarray}\label{eq:KLV}
     \mathbf{\overline V}_n&=&\mathbf{\overline V}_{n-1}-\mathbf{\dot Z}_{n-1,n}\mathbf{\overline I}_{n-1}\\
\label{eq:KLC}
     \mathbf{\overline I}_n&=&\mathbf{\overline I}_{n-1}-\mathbf{\overline I}_{n}^D \vspace{-20pt}
\end{eqnarray}
Using~(\ref{eq:KLV})--(\ref{eq:KLC}), $\mathbf{\dot S}_n=\mathbf{\overline V}_n\odot\mathbf{\check I}_n$ is rewritten as
\begin{equation}\label{eq:Sn}
     \mathbf{\dot S}_n=\mathbf{\dot S}_{n-1}-\mathbf{\dot S}_{n}^D- \mathbf{\dot Z}_{n-1,n}\,\mathbf{\overline I}_{n-1} \odot \mathbf{\check I}_{n-1}\vspace{-3pt}
\end{equation}
where $\odot$ is the Hadamard (or Schur) operator.

By straightforward algebraic manipulation of Eqs.~(\ref{eq:KLV})--(\ref{eq:Sn}) (omitted for conciseness), the forward~(\ref{eq:P_forward})--(\ref{eq:VI_forward}) and backward~(\ref{eq:P_backward})--(\ref{eq:VI_backward}) equations in the Appendix are obtained. They can be expressed in compact nonlinear form using three-phase scalar vectors as\vspace{-3pt}
\begin{eqnarray}
    \mathbf{X}_n=\mathbf{f}\left(\mathbf{X}_{n-1},\mathbf{P}^D_n,\mathbf{Q}^D_n \right)\label{eq:Xforward}\\
    \mathbf{X}_{n-1}=\mathbf{g}\left(\mathbf{X}_n,\mathbf{P}^D_n,\mathbf{Q}^D_n \right)\label{eq:Xbackward}
    \vspace{-6pt}
\end{eqnarray}
with\vspace{-6pt}
\begin{eqnarray}\label{eq:Xdefinition}
&\mathbf{X}_n=&\left[{\mathbf{P}_n}^{\!T}\,\, {\mathbf{Q}_n}^{\!T}\,\, {\mathbf{V}_{n,R}}^{\!T}\,\, {\mathbf{V}_{n,I}}^{\!T}\,\right]^T\nonumber \\[-6pt]
& &\\[-6pt]
&\mathbf{X}_{n-1}=&\left[{\mathbf{P}_{n-1}}^{\!T}\,\, {\mathbf{Q}_{n-1}}^{\!T}\,\, {\mathbf{V}_{{n-1},R}}^{\!T}\,\, {\mathbf{V}_{{n-1},I}}^{\!T}\,\right]^T\nonumber
\end{eqnarray}
\vspace{-12pt}

\subsection{Feeder model}
The feeder is composed of $N$ lines, numbered sequentially from the slack node (node \#$0$) to the ending node (node \#$N$).

Each node $n$ is described by 12 scalar variables collected in $\mathbf{X}_n$ (see Eq.~(\ref{eq:Xdefinition})), giving a total of $12 \times (N+1)$ variables for the entire problem. Lines provide $12 \times N$ equations (see~Eq. (\ref{eq:Xforward}) or~(\ref{eq:Xbackward})). To make the system well-defined, 12 additional boundary conditions are required: the slack node (\#$0$) is assigned the voltage value $\mathbf{\overline V}_{slack}$, while the ending node (\#$N$) is characterized by zero active and reactive power, that is\vspace{-3pt}
\begin{equation}\label{eq:border_feeder}
\mathbf{\overline V}_0=\mathbf{\overline V}_{slack}\, \qquad  \mathbf{\dot S}_N=\mathbf{0}
\vspace{-3pt}
\end{equation}
If the feeder is reconnected back to the slack bus to form a closed-loop, the boundary conditions become\vspace{-3pt}
\begin{equation}\label{eq:border_ring}
\mathbf{\overline V}_0=\mathbf{\overline V}_{slack} \qquad  \mathbf{\overline V}_N=\mathbf{\overline V}_{slack}
\end{equation}

\section{Solving algorithm}
\begin{figure}
    \centering
\includegraphics[width=0.55\columnwidth,height=8.7 cm]{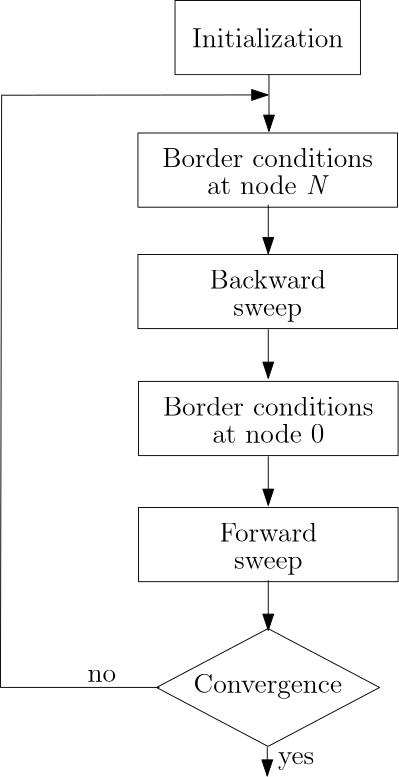}\vspace{-6pt}
    \caption{Steps of the BFS solving algorithm}
    \label{fig:BFS_algorithm}\vspace{-6pt}
\end{figure}

The solution method follows the BFS algorithm used in the classical DistFlow. The iterative procedure is described step by step with reference to the vector variables $\mathbf{X}_n$ (Fig.~\ref{fig:BFS_algorithm}). 

\subsubsection{Initialization}
The variable $\mathbf{X}_0$ is initialized by assigning initial values to the active and reactive powers $\mathbf{P}_0$ and $\mathbf{Q}_0$ (typically taken as the sum of the rated powers of all DERs and loads), while the real and imaginary components of the voltage $\mathbf{\overline V}_0$ are fixed to match those of the the slack node, according to the first of (9) or (10).
Then, all the other variables $\mathbf{X}_n$ are initialized by evaluating the forward equations~(\ref{eq:Xforward}) sequentially from $n=1$ to $n=N$.

\subsubsection{Border conditions at node \#$N$}
The update of $\mathbf{X}_N$ depends on the network topology. In the radial configuration, the real and imaginary components of the voltage remain unchanged, whereas the active and reactive powers are enforced to zero, according to the second of~(\ref{eq:border_feeder}). For the closed-loop configuration, the real and imaginary components of the voltage are enforced
to the assigned value $\mathbf{\overline V}_{slack}$, as specified by the second of (\ref{eq:border_ring}), while the apparent power is updated by adding the increment $\mathbf{\Delta \dot S}_N$, determined as follows\vspace{-3pt}
\begin{equation}\label{eq:deltaSN}
    \hspace{-0.15cm}\mathbf{\Delta \dot S}_{\!N}\!=\!\rho \mathbf{\overline V}_{\!N}\! \odot\!  \mathbf{\Delta \check J}_{\!N} \,\,\mathrm{with}\,\,\mathbf{\Delta \overline J}_N\!\!=\!\!-\mathbf{\dot Z}_{eq,N\!}^{-1}\!\left(\!\mathbf{\overline V}_{\!N}\!-\!\mathbf{\overline V}_{\!slack}\! \right)\vspace{-3pt}
\end{equation}
where the equivalent matrix $\mathbf{\dot Z}_{eq,N}$ is fixed according to the compensation method~\cite{Shirmohammadi_1988}, and the parameter $\rho$ is selected to improve convergence speed.

\subsubsection{Backward sweep}
Starting from $X_N$, the backward equations~(\ref{eq:Xbackward}) are evaluated sequentially to compute $X_{n-1}$ from $n=N$ to $n=1$.

\subsubsection{Border conditions at node \#$0$}
The real and imaginary parts of the voltage vector $\mathbf{\overline V}_0$ are set equal to the slack value $\mathbf{\overline V}_{slack}$, according to the first of~(\ref{eq:border_feeder}) or (\ref{eq:border_ring}). The update of $\mathbf{\dot S}_0$
depends on the network topology.
In the radial configuration, the active and reactive powers are left unchanged. For the closed-loop configuration, the apparent power $\mathbf{\dot S}_0$ is updated by adding the increment $\mathbf{\Delta \dot S}_0$, determined as\vspace{-3pt}
\begin{equation}\label{eq:deltaS0}
    \mathbf{\Delta \dot S}_0\!=\!\rho \mathbf{\overline V}_0\! \odot\!  \mathbf{\Delta \check J}_0 \,\,\,\mathrm{with}\,\,\mathbf{\Delta \overline J}_0\!\!=\!\mathbf{\dot Z}_{eq,0}^{-1}\!\left(\!\mathbf{\overline V}_0\!-\!\mathbf{\overline V}_{slack}\! \right)\vspace{-3pt}
\end{equation}
where the equivalent matrix $\mathbf{\dot Z}_{eq,0}$ is fixed according to the
compensation method~\cite{Shirmohammadi_1988}, and the parameter $\rho$ is the same as the one in~(\ref{eq:deltaSN}).

\subsubsection{Forward sweep}
With $\mathbf{X}_0$ determined, the forward equations~(\ref{eq:Xforward}) are solved sequentially for $n = 1, \dots, N$. Specifically, for each iteration $n$, the process begins with the voltage forward equations (\ref{eq:VR_forward})--(\ref{eq:VI_forward}), followed by the calculation of uncontrolled load power via (\ref{eq:load}), and concludes with the power forward equations (\ref{eq:P_forward})--(\ref{eq:Q_forward}).
\subsubsection{Convergence check}
The algorithm stops when the variations in $\mathbf{X}_0$ and $\mathbf{X}_N$ between successive iterations falls below a predefined tolerance; otherwise, the process returns to Step~2.\vspace{-6pt}

\section{Case Study}
\begin{figure}
    \centering \includegraphics[width=0.7\columnwidth,height=4cm]{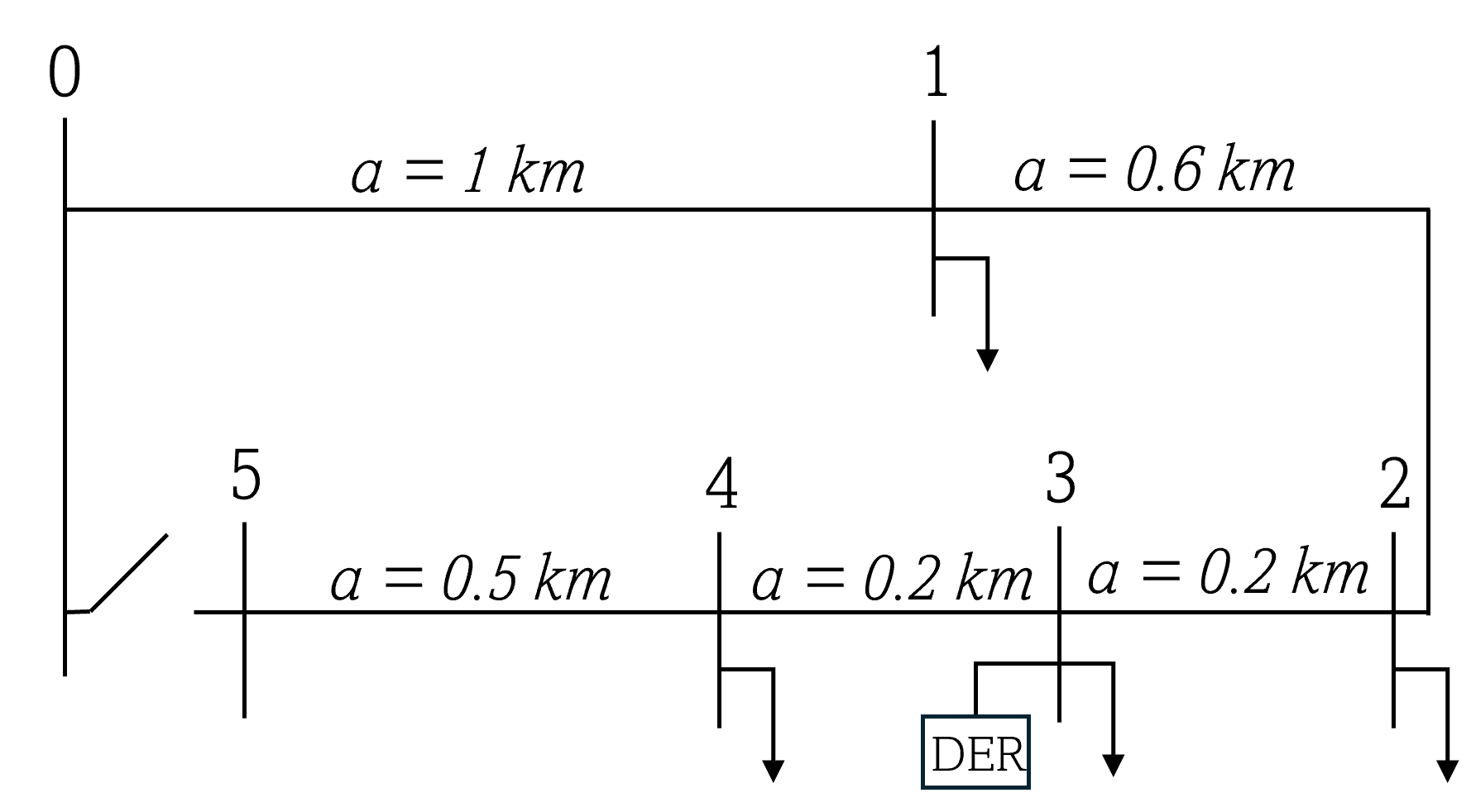}\vspace{-6pt}
    \caption{MV distribution network}
    \label{fig:rete}\vspace{-6pt}
\end{figure}

\vspace{-3pt}
The analysis considers the unbalanced three-phase MV distribution network shown in Fig.~\ref{fig:rete}, which is derived from the IEEE 13-node test feeder~\cite{testfeeders}. The line impedance matrices $\mathbf{Z}_{n-1,n}$ utilize configuration 601~\cite{testfeeders} and are expressed in $\Omega/km$, with the specific lengths of each line provided in the figure.
Node \#$0$ represents the slack bus, imposing a voltage of 4.16 kV; the feeder is converted into a closed-ring by closing the zero-impedance switch connected to node \#$5$. 
The total network load is 2.41~MW and 1.21~MVAr; it is modeled as constant power and distributed across the three phases as detailed in Table~\ref{tab:load_values}. Additionally, a balanced three-phase DER is connected at node \#$4$, providing a per-phase injection of 800~kW and 400~kVAr.
The case study first analyzes the results of the proposed
Dist3Flow under four configurations: open-ring (Cases 1 and
2) and closed-ring (Cases 3 and 4), both without and with
DERs, respectively. These results are then compared with those obtained in OpenDSS for validation. 
\begin{table}
\caption{Load parameters of the MV distribution network\vspace{-9pt}}
\begin{center}
\begin{tabular}{ccc|cc|cc}
\toprule\\[-11pt]
\multirow{3}{*}{\textbf{Node}} & 
\multicolumn{6}{c}{\textbf{3-phase load}} \\ [-2pt] 
\cmidrule{2-7}\\[-11pt]
& \makecell{$P^{L,a}$ \\ (kW)} & \makecell{$Q^{L,a}$ \\ (kVAr)} & 
  \makecell{$P^{L,b}$ \\ (kW)} & \makecell{$Q^{L,b}$ \\ (kVAr)} & 
  \makecell{$P^{L,c}$ \\ (kW)} & \makecell{$Q^{L,c}$ \\ (kVAr)} \\[-2pt] 
\midrule\\[-11pt]
1 & 517.5 & 258.8 & 258.8 & 129.4 & 258.8 & 129.4 \\
2 & 172.5 & 86.25 & 345.0 & 172.5 & 172.5 & 86.25 \\
3 & 86.25 & 43.13 & 86.25 & 43.13 & 172.5 & 86.25 \\
4 & 172.5 & 86.25 & 86.25 & 43.13 & 86.25 & 43.13 \\[-2pt]
\bottomrule
\end{tabular}\vspace{-9pt}
\label{tab:load_values}
\end{center}
\end{table}
\subsection{Analysis of the proposed model results}
\begin{figure}[t]
    \centering
    \begin{subfigure}[t]{0.49\columnwidth}
        \centering
        \includegraphics[width=\textwidth,height=10cm]{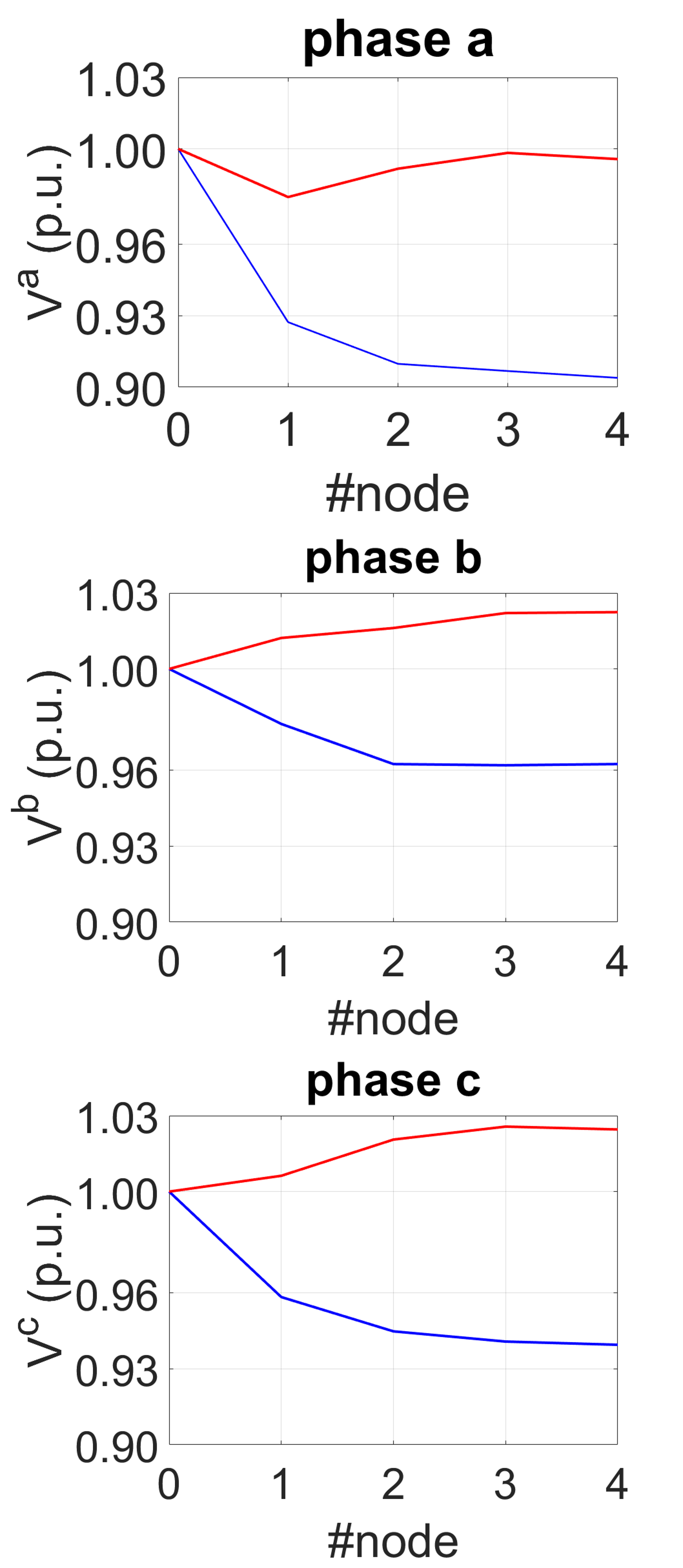}
        \caption{open ring}
        \label{fig:V_open_ring}
    \end{subfigure}
    \hfill
    \begin{subfigure}[t]{0.49\columnwidth}
        \centering
        \includegraphics[width=\textwidth,height=10cm]{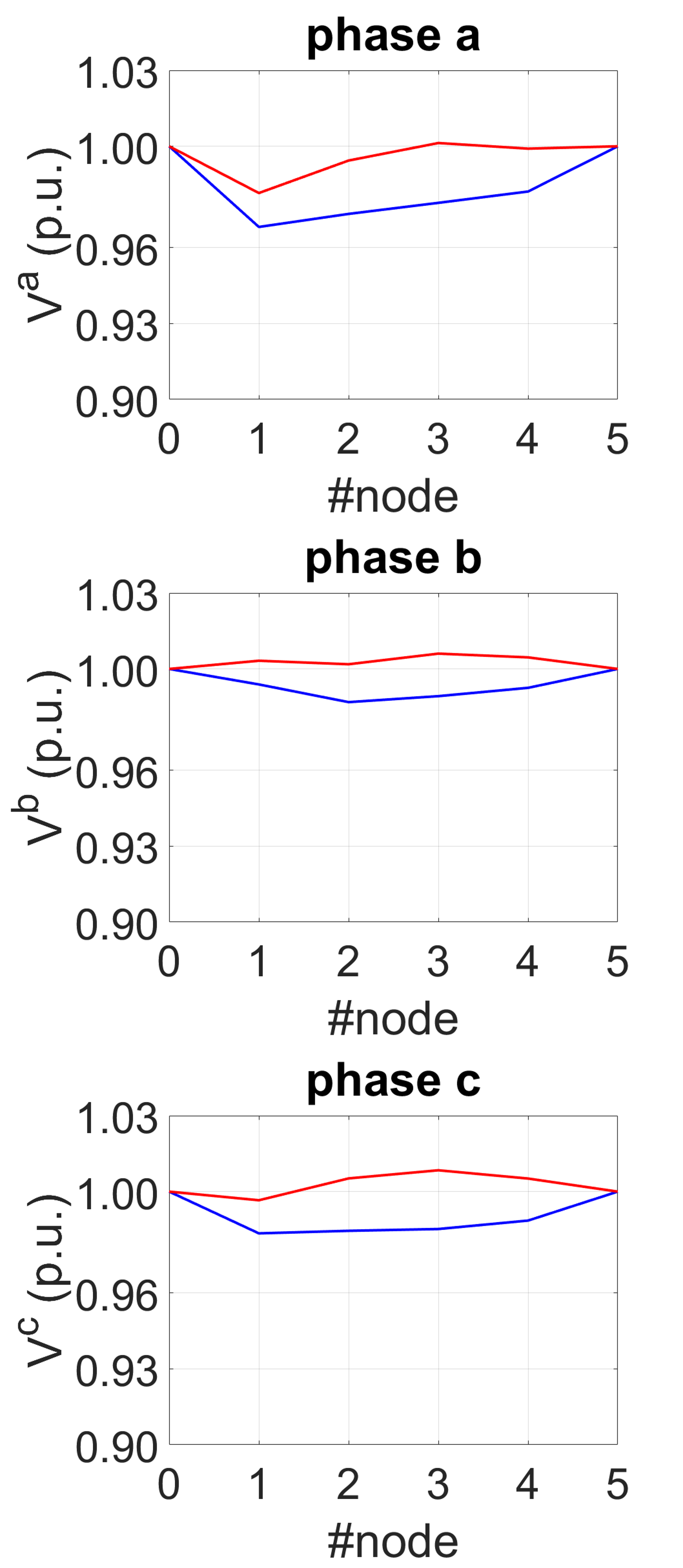}
        \caption{closed-ring}
        \label{fig:V_closed_ring}
    \end{subfigure}
    \caption{Phase voltage profiles for (a) open ring and (b) closed-ring: without DER (blue curves) and with DER (red curves)}
    \label{fig:V}
\end{figure}

\textit{Case 1}: 
Active and reactive PFs along the branches differ for each phase due to load and line imbalances, as evidenced in Table~\ref{tab:PQ_open_ring}; the most heavily loaded phase is phase $a$, with a large load at node \#$1$. Power losses are reported in Table~\ref{tab:losses_open_ring}. It can be observed that thermal power dissipation due to self-resistance (always positive) is partially offset by contributions from mutual resistance terms (which may be negative), arising from electromagnetic coupling between phases. This phenomenon can lead to negative active power losses in some branches of phase $b$. This result does not violate any physical principle: while an individual branch may exhibit incoming PF due to induction, the total three-phase system losses remain strictly positive and equal to 69.96~kW. Figure~\ref{fig:V_open_ring} shows voltage profiles that monotonically decrease, due to unidirectional PFs. However, mutual coupling determines a slight voltage rise of phase $b$ at node \#$4$ (0.9624~p.u. w.r.t. 0.9620~p.u. at node \#$3$).

\textit{Case 2}: The phase voltage profiles are shown in Figure~\ref{fig:V_open_ring} (red curves). Compared to~\textit{Case 1}, they are no longer monotonically decreasing along the feeder, due to reversals in PF caused by DER injection at node \#$3$; active and reactive branch flows are reported in Table~\ref{tab:PQ_open_ring_DER}. 
Although DER injection is balanced across the three phases, the voltage profiles differ between phases because of significant load unbalance. Overall, the presence of DERs reduces the total active power losses to 12.30~kW.

\textit{Case 3}: Closing the switch introduces PF on branch $5-0$. In this configuration, the PF direction is not uniquely determined, which may result in negative values, as reported in Table~\ref{tab:PQ_closed_ring}. Voltage profiles in Figure~\ref{fig:V_closed_ring} (blue curves) are higher than in the open-ring case (Figure~\ref{fig:V_open_ring}) and exhibit a U-shaped trend due to the ring configuration. Since both starting and ending nodes of the feeder are connected to the slack bus, the voltage at node \#$5$ is constrained to remain equal to 1~p.u, thus limiting the voltage drops and the associated losses. Then, the total active power losses amount to 21.79~kW (compared to 69.96~kW of \textit{Case 1}). 

\textit{Case 4}: The phase voltage profiles in Figure~\ref{fig:V_closed_ring} (red curves) are higher than in all previous cases. This is due to the combined effect of ring closure, which enforces equal voltages at the feeder ends and flattens the profile (especially in phase $b$), and DER injection, which further increases voltage levels through PF reversals. As shown in Table~\ref{tab:PQ_closed_ring_DER}, active and reactive branch PFs also change significantly compared to \textit{Case 3}. Overall, DERs further reduce total active power losses to 9.27~kW.

\begin{table}
\caption{Case 1: active and reactive branch powers\vspace{-9pt}}
\begin{center}
\begin{tabular}{ccc|cc|cc}
\toprule\\[-11pt]
\multirow{3}{*}{\textbf{Branch}} & 
\multicolumn{6}{c}{\textbf{3-phase active and reactive branch powers}} \\ [-2pt]
\cmidrule{2-7}\\[-11pt]
& \makecell{$P^{a}_n$ \\ (kW)} & \makecell{$Q^{a}_n$ \\ (kVAr)} & 
  \makecell{$P^{b}_n$ \\ (kW)} & \makecell{$Q^{b}_n$ \\ (kVAr)} & 
  \makecell{$P^{c}_n$ \\ (kW)} & \makecell{$Q^{c}_n$ \\ (kVAr)} \\[-2pt]
\midrule\\[-11pt]
0-1 & 999.5 & 580.0 & 771.9 & 456.7 & 713.6 & 390.9 \\
1-2 & 437.7 & 227.0 & 520.2 & 274.9 & 434.6 & 227.3 \\
2-3 & 259.4 & 131.7 & 172.3 & 86.67 & 259.6 & 130.7 \\
3-4 & 172.8 & 87.03 & 86.15 & 43.23 & 86.37 & 43.18 \\[-2pt]
\bottomrule\\[-11pt]
\end{tabular}\vspace{-9pt}
\label{tab:PQ_open_ring}
\end{center}
\end{table}

\begin{table}
\caption{Case 1: active and reactive power losses\vspace{-9pt}}
\begin{center}
\begin{tabular}{ccc|cc|cc}
\toprule
\multirow{3}{*}{\textbf{Branch}} & 
\multicolumn{6}{c}{\textbf{3-phase active and reactive power losses}} \\ [-2pt]
\cmidrule{2-7}\\[-11pt]
& \hspace{-0.2cm}$P^{a,loss}_n$ &\hspace{-0.2cm} $Q^{a,loss}_n$ &\hspace{-0.2cm} $P^{b,loss}_n$ &\hspace{-0.2cm} $Q^{b,loss}_n$ &\hspace{-0.2cm} $P^{c,loss}_n$ &\hspace{-0.2cm} $Q^{c,loss}_n$\\ 
& (kW) & \hspace{-0.2cm}(kVAr) & (kW) & \hspace{-0.2cm}(kVAr) & (kW) & \hspace{-0.2cm}(kVAr)\\ [-2pt]
\midrule\\[-11pt]
0-1 & 44.32 &\hspace{-0.2cm} 94.29 & -6.853 &\hspace{-0.2cm} 52.49 & 20.24 &\hspace{-0.2cm} 34.17 \\
1-2 & 5.827 &\hspace{-0.2cm} 9.044 & 2.635 &\hspace{-0.2cm} 15.71 & 2.491 &\hspace{-0.2cm} 10.35 \\
2-3 & 0.311 &\hspace{-0.2cm} 1.521 & -0.059 &\hspace{-0.2cm} 0.322 & 0.739 &\hspace{-0.2cm} 1.295 \\
3-4 & 0.293 &\hspace{-0.2cm} 0.778 & -0.103 &\hspace{-0.2cm} 0.102 & 0.117 &\hspace{-0.2cm} 0.057 \\[-2pt]
\bottomrule\\[-11pt]
\end{tabular}\vspace{-9pt}
\label{tab:losses_open_ring}
\end{center}
\end{table}
\begin{table}
\caption{Case 2: active and reactive branch powers\vspace{-9pt}}
\begin{center}
\begin{tabular}{ccc|cc|cc}
\toprule\\[-11pt]
\multirow{3}{*}{\textbf{Branch}} & 
\multicolumn{6}{c}{\textbf{3-phase active and reactive branch powers}} \\ [-2pt]
\cmidrule{2-7}\\[-11pt]
& \makecell{$P^{a}_n$ \\ (kW)} & \makecell{$Q^{a}_n$ \\ (kVAr)} & 
  \makecell{$P^{b}_n$ \\ (kW)} & \makecell{$Q^{b}_n$ \\ (kVAr)} & 
  \makecell{$P^{c}_n$ \\ (kW)} & \makecell{$Q^{c}_n$ \\ (kVAr)} \\ [-2pt]
\midrule\\[-11pt]
0-1 & 155.5  & 89.94  & -22.47 & -2.082 & -105.7 & -41.61 \\
1-2 & -363.9 & -172.7 & -281.5 & -131.4 & -364.3 & -173.2 \\
2-3 & -538.4 & -266.1 & -626.4 & -306.8 & -540.1 & -266.0 \\
3-4 & 172.7  & 86.89  & 86.16  & 43.22  &  86.34 & 43.17  \\  [-2pt]
\bottomrule\\[-11pt]
\end{tabular}\vspace{-6pt}
\label{tab:PQ_open_ring_DER}
\end{center}
\end{table}
\begin{table}
\caption{Case 3: active and reactive branch powers\vspace{-9pt}}
\begin{center}
\begin{tabular}{ccc|cc|cc}
\toprule
\multirow{3}{*}{\textbf{Branch}} & 
\multicolumn{6}{c}{\textbf{3-phase active and reactive branch powers}} \\ [-2pt]
\cmidrule{2-7}\\[-11pt]
& \makecell{$P^{a}_n$ \\ (kW)} & \makecell{$Q^{a}_n$ \\ (kVAr)} & 
  \makecell{$P^{b}_n$ \\ (kW)} & \makecell{$Q^{b}_n$ \\ (kVAr)} & 
  \makecell{$P^{c}_n$ \\ (kW)} & \makecell{$Q^{c}_n$ \\ (kVAr)} \\[-2pt] 
\midrule\\[-11pt]
0-1 & 439.2  & 232.2  & 319.9 & 169.5 & 285.9 & 146.6 \\
1-2 & -86.78 & -44.72 & 63.09 & 31.61 & 23.66 & 12.46 \\
2-3 & -259.4 & -131.8 & -282.3 & -141.2 & -148.7 & -73.89 \\
3-4 & -346.6  & -176.0  & -368.5  & -186.1  &  -321.3 & -160.3  \\ 
4-5 & -520.1  & -264.1  & -455.1  & -231.6  &  -408.1 & -205.1  \\
5-0 & -526.0  & -275.9  & -454.8  & -240.5  &  -410.9 & -211.1  \\[-2pt]
\bottomrule\\[-11pt]
\end{tabular}\vspace{-6pt}
\label{tab:PQ_closed_ring}
\end{center}
\end{table}
\begin{table}
\caption{Case 4: active and reactive branch powers\vspace{-9pt}}
\begin{center}
\begin{tabular}{ccc|cc|cc}
\toprule
\multirow{3}{*}{\textbf{Branch}} & 
\multicolumn{6}{c}{\textbf{3-phase active and reactive branch powers}} \\[-2pt]
\cmidrule{2-7}\\[-11pt]
& \makecell{$P^{a}_n$ \\ (kW)} & \makecell{$Q^{a}_n$ \\ (kVAr)} & 
  \makecell{$P^{b}_n$ \\ (kW)} & \makecell{$Q^{b}_n$ \\ (kVAr)} & 
  \makecell{$P^{c}_n$ \\ (kW)} & \makecell{$Q^{c}_n$ \\ (kVAr)} \\[-2pt]
\midrule\\[-11pt]
0-1 & 211.1&  112.0  & 95.90 &  51.28 & 60.21  &  30.82\\
1-2 &-308.7& -151.8  &-162.0 & -78.85 &-198.9  & -98.41\\
2-3 &-483.2& -244.2  &-506.3 & -252.2 &-372.8  & -186.0\\
3-4 & 228.1&  109.1  &207.2  &  99.82 & 254.1  &  125.9\\
4-5 & 55.41&  22.09  &120.8  & 56.11  & 167.4  &  81.60\\
5-0 & 55.47&  22.09  &120.4  & 55.61  & 167.1  & 79.96\\[-2pt]
\bottomrule\\[-11pt]
\end{tabular}\vspace{-6pt}
\label{tab:PQ_closed_ring_DER}
\end{center}
\end{table}

\subsection{Validation of the proposed model results}
To validate the proposed modeling approach, the results presented in the previous section are benchmarked against OpenDSS in terms of accuracy and computational efficiency (number of iterations). While both algorithms employ a convergence tolerance of $10^{-8}$, the comparison is conducted at a resolution of six significant figures to maintain consistency with the inherent output limitations of OpenDSS.
Table~\ref{tab:errors} summarizes the maximum absolute errors for voltage magnitude ($E_V$) and phase angle ($E_\delta$), alongside the maximum relative errors for active and reactive power flows ($\epsilon_P$ and $\epsilon_Q$). As reported, the proposed formulation yields a maximum value of $E_V=5 \times 10^{-5}$ p.u. and $E_\delta=0$ across all test cases, thereby validating the precision of the voltage calculation. Furthermore, the maximum values of $\epsilon_P$ and $\epsilon_Q$ remain well within acceptable engineering tolerances. Finally, concerning the number of iterations, both the OpenDSS and the Dist3Flow generally reach convergence in less than ten iterations, provided that an adequate value of the parameter $\rho$ in (\ref{eq:deltaS0}) and (\ref{eq:deltaSN}) is assigned (in this case it was assumed $\rho=4$).



\begin{table}[h]
\centering
\caption{Comparison with OpenDSS}
\label{tab:errors}
\begin{tabular}{c|c|c|c|c}
\toprule
\multirow{2}{*}{\textbf{Case}} & \multicolumn{2}{c|}{\textbf{\makecell{Maximum absolute error}}}  & \multicolumn{2}{c}{\textbf{\makecell{Maximum relative error}}} \\ \cline{2-5}
& \makecell{$E_V$ \\ ($10^{-5}$ p.u.)} & \makecell{$E_\delta$ \\ ($^\circ$)} & \makecell{$\epsilon_P$ \\ ($10^{-3}$ \%)} & \makecell{$\epsilon_Q$ \\ ($10^{-3}$ \%)} \\ \hline
1 & 2 & 0 & 2.7 & 4.6 \\
2 & 5 & 0 & 8.9 & 62  \\
3 & 0 & 0 & 0   & 16  \\
4 & 5 & 0 & 1.8 & 45  \\
\bottomrule
\end{tabular}
\end{table}
\vspace{-0.1cm}
\section{Conclusion}
In this paper, a backward/forward sweep (BFS) approach based on a branch flow model has been adopted for radial distribution systems with distributed energy resources. The classical DistFlow equations are extended into a three-phase formulation, termed Dist3Flow. 
By utilizing the real and imaginary components of both nodal voltages and outflowing apparent power, a complete and non-approximated formulation of the Dist3Flow equations is derived, which enables a detailed analysis of self- and mutual-coupling effects between phases in imbalanced distribution systems.
This approach enables the application of the BFS algorithm to closed-loop configurations. Comparison with OpenDSS results validates the proposed Dist3Flow equations, demonstrating high accuracy across all tested conditions. Future research will focus on extending the approach to more complex distribution systems, assessing algorithmic efficiency, and adapting these equations for use within linearized frameworks.

\begin{figure*}[!t]
\appendix\label{sec:appendix}
Forward equations, referred to phase $a$ (analog expressions are trivially derived for phases $b$ and $c$):\vspace{-12pt}
\newline
\begin{eqnarray}\label{eq:P_forward} &P_n^a&= P_{n-1}^a - P_n^{D,a} - r_{n-1,n}^{aa} \frac{{P_{n-1}^a}^2 + {Q_{n-1}^a}^2}{{V_{n-1}^a}^2}\\
&- &\hspace{-0.2cm}\sum_{i \in \{b,c\}} \frac{P_{n-1}^i P_{n-1}^a + Q_{n-1}^i Q_{n-1}^a}{{{V_{n-1}^{_i}}^2}\,\,{V_{n-1}^a}^2} \left[ r_{n-1,n}^{ai}(V_{n-1,R}^i V_{n-1,R}^a + V_{n-1,I}^i V_{n-1,I}^a) - x_{n-1,n}^{ai}(V_{n-1,I}^i V_{n-1,R}^a - V_{n-1,R}^i V_{n-1,I}^a) \right]\nonumber \\
&+ &\hspace{-0.2cm}\sum_{i \in \{b,c\}} \frac{P_{n-1}^i Q_{n-1}^a - Q_{n-1}^i P_{n-1}^a}{{{V_{n-1}^{_i}}^2}\,\,{V_{n-1}^a}^2}  \left[ r_{n-1,n}^{ai}(V_{n-1,I}^i V_{n-1,R}^a - V_{n-1,R}^i V_{n-1,I}^a) + x_{n-1,n}^{ai}(V_{n-1,R}^i V_{n-1,R}^a + V_{n-1,I}^i V_{n-1,I}^a) \right] \nonumber 
\end{eqnarray}\vspace{-12pt}
\begin{eqnarray}\label{eq:Q_forward}
&Q_n^a& = Q_{n-1}^a - Q_n^{D,a} - x_{n-1,n}^{aa} \frac{{P_{n-1}^a}^2 + {Q_{n-1}^a}^2}{{V_{n-1}^a}^2}\\
&-&\hspace{-0.2cm}\sum_{i \in \{b,c\}} \frac{P_{n-1}^i P_{n-1}^a + Q_{n-1}^i Q_{n-1}^a}{{{V_{n-1}^{_i}}^2}\,\,{V_{n-1}^a}^2} \left[ r_{n-1,n}^{ai}(V_{n-1,I}^i V_{n-1,R}^a - V_{n-1,R}^i V_{n-1,I}^a) + x_{n-1,n}^{ai}(V_{n-1,R}^i V_{n-1,R}^a + V_{n-1,I}^i V_{n-1,I}^a) \right] \nonumber\\
&-&\hspace{-0.2cm}\sum_{i \in \{b,c\}} \frac{P_{n-1}^i Q_{n-1}^a - Q_{n-1}^i P_{n-1}^a}{{{V_{n-1}^{_i}}^2}\,\,{V_{n-1}^a}^2} \left[ r_{n-1,n}^{ai}(V_{n-1,R}^i V_{n-1,R}^a + V_{n-1,I}^i V_{n-1,I}^a) - x_{n-1,n}^{ai}(V_{n-1,I}^i V_{n-1,R}^a - V_{n-1,R}^i V_{n-1,I}^a) \right] \nonumber
\end{eqnarray}\vspace{-9pt}
\begin{equation}\label{eq:VR_forward}
V_{n,R}^a = V_{n-1,R}^a - \sum_{i \in \{a,b,c\}} \left[ r_{n-1,n}^{ai} \left( \frac{P_{n-1}^i V_{n-1,R}^i + Q_{n-1}^i V_{n-1,I}^i}{{V_{n-1}^i}^2} \right) - x_{n-1,n}^{ai} \left( \frac{P_{n-1}^i V_{n-1,I}^i - Q_{n-1}^i V_{n-1,R}^i}{{V_{n-1}^i}^2} \right) \right]
\end{equation}\vspace{-9pt}
\begin{equation}\label{eq:VI_forward}
V_{n,I}^a = V_{n-1,I}^a - \sum_{i \in \{a,b,c\}} \left[ r_{n-1,n}^{ai} \left( \frac{P_{n-1}^i V_{n-1,I}^i - Q_{n-1}^i V_{n-1,R}^i}{{V_{n-1}^i}^2}\right) + x_{n-1,n}^{ai} \left( \frac{P_{n-1}^i V_{n-1,R}^i + Q_{n-1}^i V_{n-1,I}^i}{{V_{n-1}^i}^2} \right) \right]
\end{equation}
\newline\vspace{-12pt}

Backward equations, referred to phase $a$ (analog expressions are trivially derived for phases $b$ and $c$):\vspace{-12pt}
\newline
\begin{eqnarray}\label{eq:P_backward}
&P_{n-1}^a& = P_n^a  + P_n^{D,a} + r_{n-1,n}^{aa} \frac{{{P}_n^a}^2 + {{Q}_n^a}^2}{{V_{n}^a}^2}\\
&+ &\hspace{-0.2cm}\sum_{i \in \{b,c\}} \frac{{P}_n^i {P}_n^a + {Q}_n^i {Q}_n^a}{{V_n^i}^2 {V_n^a}^2} \left[ r_{n-1,n}^{ai}(V_{n,R}^i V_{n,R}^a + V_{n,I}^i V_{n,I}^a) - x_{n-1,n}^{ai}(V_{n,I}^i V_{n,R}^a - V_{n,R}^i V_{n,I}^a) \right]\nonumber \\
&- &\hspace{-0.2cm}\sum_{i \in \{b,c\}} \frac{{P}_n^i {Q}_n^a - {Q}_n^i {P}_n^a}{{V_n^i}^2 {V_n^a}^2} \left[ r_{n-1,n}^{ai}(V_{n,I}^i V_{n,R}^a - V_{n,R}^i V_{n,I}^a) + x_{n-1,n}^{ai}(V_{n,R}^i V_{n,R}^a + V_{n,I}^i V_{n,I}^a) \right] \nonumber 
\end{eqnarray}\vspace{-12pt}
\begin{eqnarray}\label{eq:Q_backward}
&Q_{n-1}^a& = Q_n^a + Q_n^{D,a} + x_{n-1,n}^{aa} \frac{{{P}_n^a}^2 + {{Q}_n^a}^2}{{V_{n}^a}^2}\\
&+&\hspace{-0.2cm}\sum_{i \in \{b,c\}} \frac{{P}_n^i {P}_n^a + {Q}_n^i {Q}_n^a}{{V_n^i}^2 {V_n^a}^2} \left[ r_{n-1,n}^{ai}(V_{n,I}^i V_{n,R}^a - V_{n,R}^i V_{n,I}^a) + x_{n-1,n}^{ai}(V_{n,R}^i V_{n,R}^a + V_{n,I}^i V_{n,I}^a) \right] \nonumber\\
&+&\hspace{-0.2cm}\sum_{i \in \{b,c\}} \frac{{P}_n^i {Q}_n^a - {Q}_n^i {P}_n^a}{{V_n^i}^2 {V_n^a}^2} \left[ r_{n-1,n}^{ai}(V_{n,R}^i V_{n,R}^a + V_{n,I}^i V_{n,I}^a) - x_{n-1,n}^{ai}(V_{n,I}^i V_{n,R}^a - V_{n,R}^i V_{n,I}^a) \right] \nonumber
\end{eqnarray}\vspace{-9pt}
\begin{equation}\label{eq:VR_backward}
V_{n-1,R}^a = V_{n,R}^a + \sum_{i \in \{a,b,c\}} \left[ r_{n-1,n}^{ai} \left( \frac{{P}_n^i V_{n,R}^i + {Q}_n^i V_{n,I}^i}{{V_n^i}^2} \right) - x_{n-1,n}^{ai} \left( \frac{{P}_n^i V_{n,I}^i - {Q}_n^i V_{n,R}^i}{{V_n^i}^2} \right) \right]
\end{equation}\vspace{-9pt}
\begin{equation}\label{eq:VI_backward}
V_{n-1,I}^a = V_{n,I}^a + \sum_{i \in \{a,b,c\}} \left[ r_{n-1,n}^{ai} \left( \frac{{P}_n^i V_{n,I}^i - {Q}_n^i V_{n,R}^i}{{V_n^i}^2} \right) + x_{n-1,n}^{ai} \left( \frac{{P}_n^i V_{n,R}^i + {Q}_n^i V_{n,I}^i}{{V_n^i}^2} \right) \right]
\end{equation}\vspace{-12pt}
\end{figure*}

\bibliographystyle{IEEEtran}
\bibliography{Bibliography.bib}

\end{document}